\documentclass[11pt]{article}

\usepackage[preprint]{acl}

\usepackage{times}
\usepackage{latexsym}

\usepackage[T1]{fontenc}
\usepackage[utf8]{inputenc}
\usepackage{microtype}
\usepackage{inconsolata}
\usepackage{multirow}
\usepackage{booktabs}
\usepackage{geometry}
\usepackage{graphicx}
\usepackage{amsmath}
\usepackage{makecell}
\usepackage{amsfonts}
\usepackage{utfsym}
\usepackage{arydshln}
\usepackage[most]{tcolorbox}
\usepackage[normalem]{ulem}
\usepackage{xcolor} 
\usepackage{float}
\usepackage{enumitem}

\title{SAC: Neural Speech Codec with Semantic-Acoustic \\ Dual-Stream Quantization}

\author{
 \textbf{Wenxi Chen\textsuperscript{1,2}\thanks{Work done during an internship at Soul AI Lab.}},
 \textbf{Xinsheng Wang\textsuperscript{3}\thanks{Corresponding author.}},
 \textbf{Ruiqi Yan\textsuperscript{1}},
 \textbf{Yushen Chen\textsuperscript{1,2}},
 \textbf{Zhikang Niu\textsuperscript{1,2}},
 \textbf{Ziyang Ma\textsuperscript{1}},
 \\
 \textbf{Xiquan Li\textsuperscript{1}},
 \textbf{Yuzhe Liang\textsuperscript{1,2}},
 \textbf{Hanlin Wen\textsuperscript{3}},
 \textbf{Shunshun Yin\textsuperscript{3}},
 \textbf{Ming Tao\textsuperscript{3}},
 \textbf{Xie Chen\textsuperscript{1,2}\footnotemark[2]}
\\
\\
 \textsuperscript{1}X-LANCE Lab, Shanghai Jiao Tong University, China
\\
 \textsuperscript{2}Shanghai Innovation Institute, China
 \textsuperscript{3}Soul AI Lab, China
\\
\small{
  \href{mailto:1029713857@sjtu.edu.cn}{1029713857@sjtu.edu.cn};
  \href{mailto:wangxinsheng@soulapp.cn}{wangxinsheng@soulapp.cn};
  \href{mailto:chenxie95@sjtu.edu.cn}{chenxie95@sjtu.edu.cn}
}
}

\begin{document}
\maketitle

\begin{abstract}
Speech codecs that convert continuous speech signals into discrete tokens have become essential for speech language models. However, existing codecs struggle to balance high-quality reconstruction with semantically rich representations, limiting their effectiveness in both generative and understanding tasks.  
In this work, we propose \textbf{SAC}, a neural speech codec with semantic-acoustic dual-stream quantization. By disentangling semantic and acoustic modeling into two dedicated streams, SAC enables each to be optimized for its respective role. 
Comprehensive evaluations show that SAC achieves strong reconstruction performance across diverse bitrates under both clean and noisy conditions, with particularly high scores on UTMOS and WER, indicating superior naturalness and intelligibility. Moreover, SAC substantially surpasses prior codecs in semantic representation, approaching the level of continuous self-supervised embeddings. 
When used as a tokenizer for LLM-based text-to-speech, SAC enables a single-stage autoregressive (AR) TTS model that clearly outperforms state-of-the-art AR systems.
Our disentanglement analysis further validates the effectiveness of the dual-stream design, offering new potential for controllable speech generation.
The code and pre-trained models are available at \url{https://github.com/Soul-AILab/SAC}.
\footnote{Demo at \url{https://sac-codec.github.io}.}
\end{abstract}

\section{Introduction}
With the rapid advancement of large language models (LLMs), speech language models (SLMs) have emerged by extending text-based LLMs with speech modalities~\citep{survey-pengjing,wavchat}. Central to these models is the speech tokenizer, which discretizes continuous speech waveforms into token sequences, thereby enabling seamless integration with token-based language models~\citep{review-yiwei}. Leveraging such speech tokens, SLMs have driven progress in a wide range of downstream applications, including text-to-speech (TTS)~\citep{fireredtts2,spark-tts}, speech understanding~\citep{token-understanding}, and spoken dialogue systems ~\citep{Qwen3-Omni,glm-4-voice,slam-omni,moshi}.

\begin{figure}[t]
    \centering
    \includegraphics[width=0.48\textwidth]{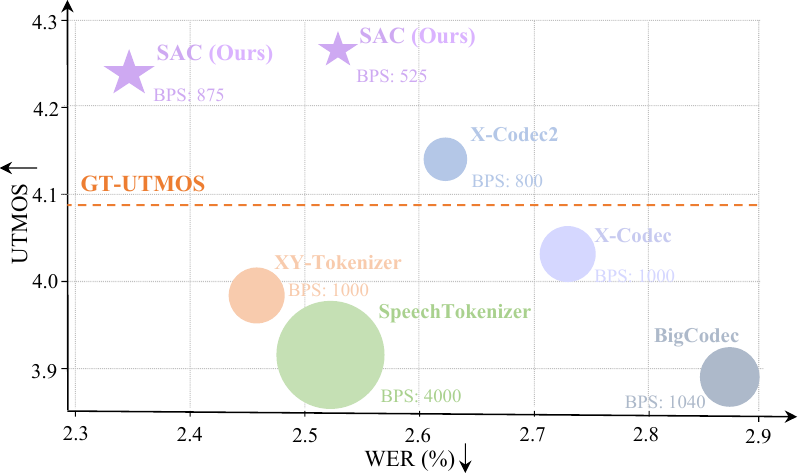}
    \caption{Comparison of codecs on speech reconstruction. The x-axis shows WER, reflecting speech intelligibility, while the y-axis presents UTMOS, reflecting objective naturalness. Circle size indicates bitrate.}
    \label{fig:Codec_compare}
    \vspace{-4mm}
\end{figure}

Semantic tokens are among the most widely used tokens in speech processing. These tokens are typically derived from either self-supervised models~\citep{case-study-yifan} or supervised models~\citep{Cosyvoice, Cosyvoice2}, making them effective in capturing semantic meaning. However, the absence of essential acoustic information significantly limits the applicability of semantic tokens. In contrast, acoustic tokens are usually generated by neural audio codecs trained with reconstruction objectives~\citep{encodec, Soundstream}. While this approach preserves fine-grained acoustic details, the lack of semantic supervision results in weaker alignment with semantic content and reduced compatibility with text-based LMs~\citep{AudioCodecBench, CodecBench}.

To enhance semantic representation, recent advancements in speech codecs have explored incorporating semantic supervision during training.
For example, SpeechTokenizer~\citep{speechtokenizer} employs semantic distillation, where representations from an SSL model guide the output of the first residual vector quantization (RVQ) layer. The X-Codec series~\citep{xcodec,llasa-xcodec2} and XY-Tokenizer~\citep{XY-Tokenizer} adopt an “X-shaped” paradigm, explicitly injecting pre-trained semantic features and fusing them with acoustic embeddings before quantization. 
However, while these approaches improve semantic alignment compared to methods without semantic constraints, they still fall short of pure semantic tokens in terms of semantic relevance.
This raises a central question: \textit{Can semantic and acoustic tokens be disentangled at the token level, allowing each to specialize in its respective role?} In this work, we explore such a dual-stream design that decouples semantic and acoustic modeling into independent streams, and show that it leads to improved semantic representation and reconstruction performance.

In this paper, we propose \textbf{SAC}, a novel \textbf{S}emantic\\–\textbf{A}coustic Dual-Stream Neural Speech \textbf{C}odec. Unlike prior approaches that inject semantic supervision into codecs, 
SAC complements the semantic tokens by introducing a separate acoustic token stream, which provides the essential acoustic information missing from the semantic tokens, all while ensuring the integrity of the semantic representations.
Specifically, in the semantic stream, we adopt a pre-trained speech tokenizer~\citep{glm4voice-tokenizer} to extract semantic tokens aligned with linguistic content, keeping it frozen during training to ensure faithful retention of semantic information. In the acoustic stream, we follow the design of neural audio codecs~\citep{dac}, employing temporally distributed acoustic tokens to capture the essential acoustic information, e.g., timbre and emotional attributes, that is missing from the semantic tokens.
This dual-stream design unifies the complementary strengths of both within a single framework: semantic tokenizers excel in speech understanding and dialogue tasks~\citep{glm-4-voice,kimi-audio}, while acoustic tokenizers are particularly effective in generative modeling~\citep{valle,valle2}. 
At the decoding stage, a ConvNeXt-based~\citep{convnext} prenet is employed to fuse the two streams of embeddings, followed by a codec decoder that reconstructs the waveform.  
% During codec training, we further incorporate auxiliary supervision: (i) semantic feature supervision based on Whisper~\citep{whisper} to emphasize linguistic fidelity, and (ii) speaker feature supervision based on ERes2Net~\citep{eres2net-sv} to enhance global timbre modeling. 
Experiments show that SAC delivers strong reconstruction and semantic relevance, while also supporting competitive performance in downstream LLM-based TTS tasks.
Our main contributions can be summarized as follows:

% \vspace{-1mm}
\begin{itemize}[leftmargin=*,noitemsep]
    \item We propose \textbf{SAC}, a semantic–acoustic dual-stream neural speech codec. By explicitly disentangling speech encoding into parallel semantic and acoustic streams, SAC enables each pathway to specialize in modeling linguistic content and acoustic detail, respectively. 
    % \item We incorporate speaker feature supervision into codec training, which substantially improves the preservation of timbre information while incurring negligible cost in reconstruction quality.  
    \item SAC outperforms existing codecs with strong speech reconstruction quality and semantic representation across different bitrates. 
    \item We validate SAC in TTS downstream tasks, where SAC-based single-stage AR models significantly surpass SOTA pure AR systems in both intelligibility and objective naturalness.
    \item We further analyze SAC’s effectiveness in speech disentanglement, paving the way for controllable and anonymized speech applications.
\end{itemize}

% To support future research, we will release the full codebase and pre-trained SAC models.  

\begin{figure*}[t]
    \centering
    \includegraphics[width=\textwidth]{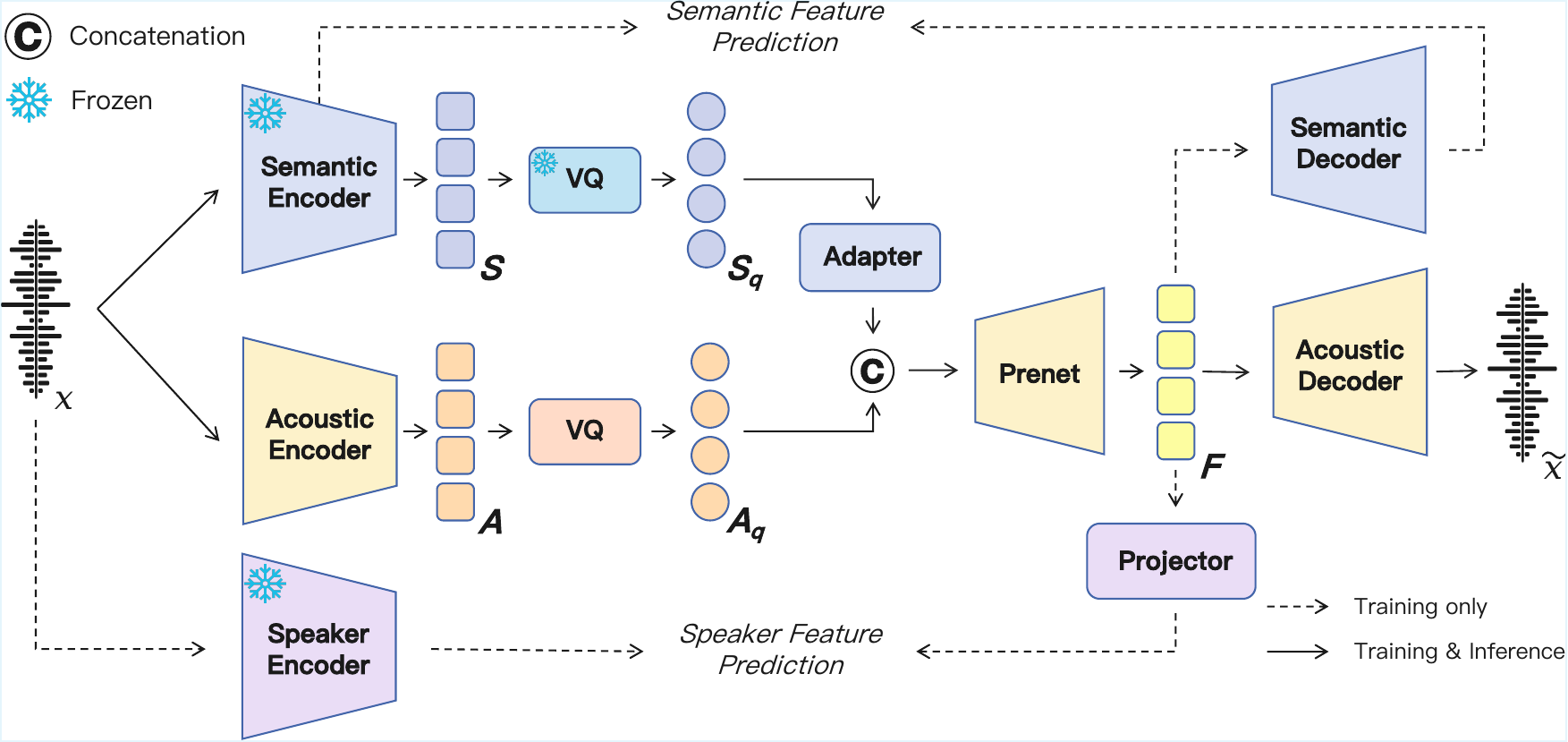}
    \caption{
    Overview of SAC. Semantic and speaker feature supervision are applied only during codec training, with their respective encoders kept frozen to preserve the integrity of extracted features.
    }
    \label{fig:SAC}
    % \vspace{-2mm}
\end{figure*}

\section{Related Work}

% \vspace{-2mm}
\paragraph{Semantic Tokenizer}
Semantic tokens are derived from quantized representations of self-supervised learning (SSL) models or supervised models. They are strongly correlated with textual semantics but are not primarily designed for waveform reconstruction.
Following the introduction of the $\mathcal{S}^3$ tokenizer in CosyVoice~\citep{Cosyvoice}, a range of supervised semantic tokenizers have been proposed.  For example, CosyVoice3 \citep{Cosyvoice3} employs multi-task learning to improve prosody modeling in their tokenizer, while the Baichuan Audio Tokenizer \citep{Baichuan-audio} introduced an additional mel reconstruction objective to capture both semantic and acoustic details. 
However, due to the lack of necessary acoustic information, the application of semantic tokens in generation tasks still requires additional generative models and reference audio to generate acoustic features.
To fully leverage the semantic consistency inherent in semantic tokens while addressing their lack of acoustic information during generation tasks, we propose a method that preserves their semantic integrity by introducing a separate acoustic token stream.

\paragraph{Neural Audio Codecs}
Acoustic tokens are derived from neural audio codecs, which compress continuous audio signals into discrete tokens while reconstructing high-fidelity waveforms~\citep{NDVQ}. Speech codecs typically adopt the VQ-GAN paradigm~\citep{vqgan}, where a VQ-VAE~\citep{vqvae}-based generator performs the encode–quantize–decode process, and adversarial discriminators distinguish real from synthetic speech to improve perceptual quality.  
SoundStream~\citep{Soundstream} introduced residual vector quantization (RVQ) to improve quantization efficiency, while Encodec~\citep{encodec} incorporated LSTMs into the encoder–decoder structure to enhance compression.
More recently, several single-codebook codecs have emerged, including Single-Codec~\citep{Single-codec}, WavTokenizer~\citep{wavtokenizer}, and BigCodec~\citep{bigcodec}, which replace multi-layer quantizers with a single codebook. These designs achieve ultra-low-bitrate compression while simplifying and accelerating downstream modeling.  
However, the absence of explicit semantic consistency constraints limits their effectiveness in recognition-oriented tasks and hinders compatibility with text-based LMs~\citep{AudioCodecBench}.

% \vspace{-2mm}
\paragraph{Codec with Semantic Supervision}
To enhance semantic representation, recent speech codecs have introduced semantic features through distillation or injection. For instance, SpeechTokenizer \citep{speechtokenizer} leverages HuBERT \citep{hubert} to guide the first RVQ layer toward encoding semantic information, while Mimi \citep{moshi} distills WavLM \citep{wavlm} features into a separate VQ module. X-Codec \citep{xcodec,llasa-xcodec2} introduces semantic injection via an auxiliary semantic module, and XY-Tokenizer \citep{XY-Tokenizer} further enhances textual alignment through an LLM-based ASR objective.
However, these methods rely on fused semantic–acoustic tokens that must jointly support semantic prediction and spectrogram reconstruction, limiting both semantic fidelity and reconstruction quality. Although SemantiCodec \citep{semanticodec}, built on AudioMAE \citep{audiomae}, seeks to disentangle semantic and acoustic streams, our analysis indicates that the separation remains incomplete. In contrast, SAC employs a dual-stream semantic–acoustic design, enabling both stronger semantic representation and improved reconstruction.

\section{SAC}
To jointly leverage the semantic modeling capabilities of speech tokenizers and the fine-grained acoustic representations of neural codecs, we propose the semantic-acoustic dual-stream codec (SAC). As illustrated in Fig.~\ref{fig:SAC}, SAC employs two discrete encoding streams: (1) a \textit{semantic stream}, which utilizes a pre-trained semantic tokenizer to model linguistic content, and 
(2) an acoustic stream, which relies on a speech codec to provide the acoustic information that is missing from the semantic tokens.
Together, these two streams enable a more comprehensive representation of speech signals and explicitly mitigate conflicts between speech tokens when optimizing for these two distinct objectives during codec training.

\subsection{Model Architecture}
SAC is built on the VQ-GAN framework~\citep{vqgan}, which follows a VQ-VAE architecture~\citep{vqvae-speech} to reconstruct raw speech. As shown in Fig.~\ref{fig:SAC}, SAC comprises a dual-stream encoder–quantizer and a unified codec decoder, complemented by auxiliary modules for semantic and speaker feature supervision. The details of each component are described in the following subsections.

% \vspace{-1mm}
\paragraph{Semantic Stream} 
To ensure that the semantic stream of SAC maintains strong semantic consistency, we employ a pre-trained semantic tokenizer, keeping its parameters frozen throughout SAC training. Specifically, we use the speech tokenizer proposed in~\citet{glm4voice-tokenizer}, which tokenizes input speech into discrete tokens at a frame rate of 12.5 Hz.
Formally, given an input waveform $x$, the semantic tokenizer first extracts fine-grained continuous representations $\mathbf{S}_c$ at 50 Hz, which also serve as the target for auxiliary semantic supervision. A temporal pooling layer then downsamples $\mathbf{S}_c$ to 12.5 Hz, producing $\mathbf{S}$. These features are quantized via a vector quantization layer to obtain discrete semantic tokens and their corresponding quantized embeddings $\mathbf{S}_q$. 
During training, the semantic tokenizer is kept frozen to ensure that the semantic stream focuses exclusively on linguistic content without being biased toward acoustic details. 
To achieve temporal alignment with the acoustic embeddings, $\mathbf{S}_q$ is further upsampled by a ConvNeXt-based adapter~\citep{convnext}, resulting in the semantic features $\mathbf{S}_q'$.

% \vspace{-1mm}
\paragraph{Acoustic Stream}

The acoustic token serves to complement the acoustic details missing from the semantic token. 
Specifically, we adopt the Encodec architecture~\citep{encodec}, which employs stacked convolutional and temporal downsampling layers with stride $\tau$ to extract frame-level acoustic representations $\mathbf{A}$. Following DAC~\citep{dac}, we apply factorized code projections to map $\mathbf{A}$ into a lower-dimensional embedding space and perform single-codebook quantization based on $L_2$ distances. 
To mitigate codebook underutilization, entries that remain inactive over prolonged training intervals are reinitialized with randomly sampled embeddings drawn from the current batch~\citep{Jukebox}.
Since acoustic information is inherently more fine-grained than semantic information, we adopt higher frame rates for its representation, specifically 25 Hz for low-bitrate settings and 50 Hz for high-bitrate settings.
The corresponding strides $\tau$ are set to $(2, 2, 4, 5, 8)$ and $(2, 4, 5, 8)$, respectively, yielding temporal reduction factors of 640 and 320 for input audio sampled at 16 kHz.

% \vspace{-1mm}
\paragraph{Decoder} 
The quantized acoustic embeddings $\mathbf{A}_q$ are concatenated with the semantic embeddings $\mathbf{S}_q'$ along the feature dimension to form a unified representation $\mathbf{U}$. This joint representation is then processed by a ConvNeXt-based prenet, which upsamples it to 50 Hz, producing the fused feature sequence $\mathbf{F}$. The fused representation $\mathbf{F}$ integrates linguistic information from the semantic stream with timbre and acoustic detail from the acoustic stream. Subsequently, $\mathbf{F}$ is passed through a mirrored decoder composed of stacked convolutional and temporal upsampling layers to reconstruct the waveform $\tilde{x}$, with deconvolution strides set to $\tau = (8, 5, 4, 2)$.  
Following the design of “X-shaped” codec models~\citep{xcodec}, we introduce an auxiliary semantic reconstruction objective to ensure that key linguistic information is preserved during decoding. Specifically, $\mathbf{F}$ is fed into a CNN-based semantic decoder to predict the reconstructed semantic features $\tilde{\mathbf{S}}_c$. A mean squared error (MSE) loss,  
\begin{equation}
\mathcal{L}_{\text{sem}} = \|\tilde{\mathbf{S}}_c - \mathbf{S}_c\|_2^2
\label{eq:sem-loss}
\end{equation}
is then applied between $\tilde{\mathbf{S}}_c$ and the ground-truth semantic features $\mathbf{S}_c$ to regularize training.

\subsection{Auxiliary Speaker Feature Supervision}
While the acoustic stream effectively captures fine-grained spectral details, it may insufficiently model global timbre characteristics. To mitigate this limitation and improve timbre reconstruction, we introduce explicit speaker feature supervision.
To be specific, we employ ERes2Net~\citep{eres2net-sv} to extract speaker embeddings $\mathbf{S}_p$ as the supervision target. 
%ERes2Net has demonstrated strong performance in speaker verification and, due to its architectural distinction from WavLM~\citep{wavlm-sv} used for similarity evaluation, helps prevent feature overfitting.  
For prediction, we compute the temporal mean and variance of the fused representation $\mathbf{F}$ and concatenate them into a global feature $\mathbf{f}$. This vector is passed through a lightweight two-layer MLP projector to generate the predicted speaker embedding $\tilde{\mathbf{S}}_p$. An MSE loss is then applied between $\tilde{\mathbf{S}}_p$ and $\mathbf{S}_p$ to encourage accurate modeling of timbre information:

\vspace{-4mm}
\begin{align}
\mathbf{f} &= [\,\text{Mean}_t(\mathbf{F});\;\text{Std}_t(\mathbf{F})\,], \\
\mathcal{L}_{\text{spk}} &= \left\|\tilde{\mathbf{S}}_p - \mathbf{S}_p\right\|_2^2
= \Big\|\text{Proj}(\mathbf{f}) - \mathbf{S}_p\Big\|_2^2.
\end{align}

\subsection{Training Objectives}
SAC is optimized under the VQ-GAN framework, where the overall objective comprises losses for both the generator and the discriminator.

\paragraph{Reconstruction Loss}
Following DAC, we define the reconstruction loss $\mathcal{L}_{\text{recon}}$ as the $L_1$ distance between the reconstructed and ground-truth audio signals across multiple scales, applied on both log-scale and linear-scale spectrograms.

% \vspace{-1mm}
\paragraph{VQ Loss} 
For the acoustic stream, the codebook is optimized by minimizing the $L_2$ distance between the encoder outputs and their quantized embeddings, with gradients propagated using the straight-through estimator (STE)~\citep{ste}. The VQ loss $\mathcal{L}_{\text{vq}}$ also incorporates a commitment term that constrains encoder outputs to remain close to their assigned codebook entries.

% \vspace{-1mm}
\paragraph{Discriminative Loss} 
We employ a multi-period discriminator (MPD)~\citep{hifi-gan} and a multi-scale STFT-based discriminator (MS-STFT)~\citep{encodec}, following~\citet{bigcodec}. The discriminators are optimized using the least-squares GAN objective~\citep{lsgan}. For the generator, we apply both an adversarial loss $\mathcal{L}_{\text{adv}}$ and a feature matching loss $\mathcal{L}_{\text{feat}}$, the latter computed as the $L_1$ distance between intermediate feature maps of real and generated audio.  

The overall generator loss is formulated as:  
\vspace{-2mm}
\begin{align}
\mathcal{L}_G = \;&
\lambda_{\text{recon}} \mathcal{L}_{\text{recon}} +
\lambda_{\text{vq}} \mathcal{L}_{\text{vq}} +
\lambda_{\text{adv}} \mathcal{L}_{\text{adv}} \nonumber \\
&+
\lambda_{\text{feat}} \mathcal{L}_{\text{feat}} +
\lambda_{\text{sem}} \mathcal{L}_{\text{sem}} +
\lambda_{\text{spk}} \mathcal{L}_{\text{spk}},
\label{eq:gen-loss}
\end{align}
where each coefficient $\lambda$ is a tunable hyperparameter weighting the corresponding objective.

\section{Experimental Setup}
\subsection{Training Details}

\paragraph{Datasets}
To ensure diversity in the training data, we randomly sampled approximately 20,000 hours of bilingual (Chinese and English) speech data from various sources. These include Emilia~\citep{emilia}, WenetSpeech4TTS~\citep{wenetspeech4tts}, LibriSpeech~\citep{librispeech}, Libriheavy~\citep{libriheavy}, MLS~\citep{mls}, and in-house data. Further details of the training data are provided in Appendix~\ref{app:SAC_data}.

\paragraph{Training Setup}
Both the semantic and acoustic codebooks in SAC contain 16,384 entries. 
%To examine performance at different bitrates, 
To provide configurations for different bitrates, we set the acoustic token frame rate to 25 Hz or 50 Hz, corresponding to overall token rates of 37.5 Hz and 62.5 Hz, respectively. Models are trained for 850k steps on 8 NVIDIA H20 GPUs with a batch size of 24. During training, each audio sample is randomly cropped into 2.4-second segments. 
Optimization is performed using the AdamW optimizer~\citep{adamw} with $\beta_1 = 0.8$ and $\beta_2 = 0.9$. Both the generator and discriminator learning rates are initialized at $1 \times 10^{-4}$ and decayed exponentially throughout training. Additional training details are provided in Appendix~\ref{app:SAC_train_details}.

\subsection{Evaluation Details}
An ideal speech token should not only have strong audio reconstruction capability but also maintain good semantic consistency, facilitating tasks such as audio generation or comprehension. Therefore, we conduct a comprehensive evaluation of SAC from both reconstruction and semantic representation perspectives.

\paragraph{Speech Reconstruction}
We evaluate speech reconstruction performance on the LibriSpeech \textit{test-clean} set, which contains 2,620 utterances at 16 kHz. 
%For computational efficiency, 
To facilitate comparison across models, we report key parameters, including codebook size, the number of quantizers (Nq), token rate, and bandwidth (BPS). Speech intelligibility is assessed using Short-Time Objective Intelligibility (STOI) and Word Error Rate (WER), where transcriptions are obtained with the HuBERT-based~\citep{hubert} model.
% \footnote{\url{https://huggingface.co/facebook/hubert-large-ls960-ft}}. 
Acoustic quality is measured by the Perceptual Evaluation of Speech Quality (PESQ) and UTMOS~\citep{utmos}, while speaker similarity (SIM) is computed via a WavLM-based~\citep{wavlm-sv} speaker verification model.
% \footnote{\url{https://github.com/microsoft/UniSpeech/tree/main/downstreams/speaker_verification}}. 
For comparison, we evaluate SAC at two token rates against a range of state-of-the-art codecs with similar bitrates, including DAC~\citep{dac}, Encodec~\citep{encodec}, Mimi~\citep{moshi}, SpeechTokenizer~\citep{speechtokenizer}, SemantiCodec~\citep{semanticodec}, BigCodec~\citep{bigcodec}, the X-Codec series~\citep{xcodec,llasa-xcodec2}, XY-Tokenizer~\citep{XY-Tokenizer}, WavTokenizer~\citep{wavtokenizer}, MagiCodec~\citep{magicodec}, and TS3-Codec~\citep{Ts3-codec}. All baselines are reproduced on our test set using their official checkpoints. For ablations, we consistently adopt the lower-bitrate configuration of SAC for comparison.

\begin{table*}[t]
\centering
\small
\resizebox{1\linewidth}{!}{
\begin{tabular}{l!{\vrule width 0.8pt}cccc!{\vrule width 0.8pt}ccccccc}
\toprule
Model & \makecell{Codebook \\ Size} & Nq & \makecell{Token \\ Rate} & BPS & STOI$\uparrow$ & \makecell{PESQ \\ NB$\uparrow$} & \makecell{PESQ \\ WB$\uparrow$} & UTMOS$\uparrow$ & SIM$\uparrow$ & WER(\%)$\downarrow$ \\
\midrule
Ground Truth & - & - & - & - & 1.00 & 4.55 & 4.64 & 4.09 & 1.00 & 2.16 \\
\midrule
\midrule
DAC & 1024 & 12 & 600 & 6000 & \textbf{0.97} & \textbf{4.15} & \textbf{4.01} & \textbf{4.00} & \textbf{0.95} & \textbf{2.22} \\
Encodec & 1024 & 8 & 600 & 6000 & 0.94 & 3.18 & 2.77 & 3.09 & 0.89 & 2.36 \\
Mimi & 2048 & 32 & 400 & 4400 & 0.96 & 3.80 & 3.45 & 3.95 & 0.93 & 2.27 \\
SpeechTokenizer & 1024 & 8 & 400 & 4000 & 0.92 & 3.05 & 2.60 & 3.90 & 0.85 & 2.51 \\
\midrule
\midrule
DAC & 1024 & 3 & 150 & 1500 & 0.79 & 1.61 & 1.25 & 1.48 & 0.47 & 7.80 \\ 
Encodec & 1024 & 2 & 150 & 1500 & 0.85 & 1.94 & 1.56 & 1.58 & 0.60 & 5.62 \\
SemantiCodec & ~32768/8192$^\ddagger$ & ~1/1$^\ddagger$ & 100 & 1400 & 0.88 & 2.63 & 2.02 & 2.94 & 0.72 & 3.31 \\
Mimi & 2048 & 8 & 100 & 1100 & 0.91 & 2.80 & 2.26 & 3.63 & 0.74 & 3.24 \\
BigCodec & 8192 & 1 & 80 & 1040 & \textbf{0.94} & \textbf{3.27} & \textbf{2.68} & \textbf{4.11} & \textbf{0.84} & 2.92 \\
% DAC & 1024 & 2 & 100 & 1000 & 0.73 & 1.40 & 1.13 & 1.29 & 0.32 & 23.14 \\ % bad
X-codec & 1024 & 2 & 100 & 1000 & 0.86 & 2.68 & 2.11 & 4.06 & 0.68 & 2.73 \\
XY-Tokenizer & 1024 & 8 & 100 & 1000 & 0.91 & 3.00 & 2.41 & 3.98 & \textbf{0.84} & \textbf{2.46} \\
\midrule
\midrule
WavTokenizer & 4096 & 1 & 75 & 900 & 0.90 & 2.63 & 2.13 & 3.79 & 0.65 & 4.15 \\
MagiCodec & 131072 & 1 & 50 & 850 & 0.92 & \textbf{3.16} & 2.54 & 4.17 & 0.77 & 3.52 \\
TS3-Codec$^\dagger$ & 131072 & 1 & 50 & 850 & 0.91 & - & 2.23 & 3.84 & 0.68 & 3.60 \\
X-codec2 & 65536 & 1 & 50 & 800 & 0.92 & 3.04 & 2.43 & 4.13 & 0.82 & 2.61 \\
\midrule
SAC (ours) & ~16384/16384$^\ddagger$ & ~1/1$^\ddagger$ & 62.5 & 875 & \textbf{0.93} & 3.15 & \textbf{2.59} & \textbf{4.25} & \textbf{0.86} & \textbf{2.35} \\
\bottomrule
\end{tabular}
}
\caption{Comparison of high-bitrate codec models on speech reconstruction metrics. \textbf{Bold} numbers denote the best performance among models with comparable bitrates. $^\dagger$TS3-Codec results are taken from the original paper. $^\ddagger$For codec models with semantic–acoustic decoupling (e.g., SAC and SemantiCodec), the codebook size and Nq are reported as ``\textit{x}/\textit{y}'', where \textit{x} corresponds to the semantic stream and \textit{y} to the acoustic stream.}
\label{tab:codec-high-recon}
\end{table*}

\begin{table*}[htbp]
\centering
\small
\resizebox{1\linewidth}{!}{
\begin{tabular}{l!{\vrule width 0.8pt}cccc!{\vrule width 0.8pt}ccccccc}
\toprule
Model & \makecell{Codebook \\ Size} & Nq & \makecell{Token \\ Rate} & BPS & STOI$\uparrow$ & \makecell{PESQ \\ NB$\uparrow$} & \makecell{PESQ \\ WB$\uparrow$} & UTMOS$\uparrow$ & SIM$\uparrow$ & WER(\%)$\downarrow$ \\
\midrule
Ground Truth & - & - & - & - & 1.00 & 4.55 & 4.64 & 4.09 & 1.00 & 2.16 \\
\midrule
\midrule
Encodec & 1024 & 1 & 75 & 750 & 0.77 & 1.48 & 1.23 & 1.25 & 0.25 & 41.2 \\
SemantiCodec & 32768/8192 & 1/1 & 50 & 700 & 0.86 & 2.33 & 1.78 & 2.94 & 0.61 & 5.54 \\
% Mimi & 2048 & 5 & 62.5 & 687.5 & 0.87 & 2.33 & 1.84 & 3.29 & 0.58 & 4.31 \\
TS3-Codec & 131072 & 1 & 40 & 680 & \textbf{0.90} & - & 2.06 & 3.73 & 0.63 & 4.50 \\
% DAC & 1024 & 1 & 50 & 500 &  &  &  &  &  &  \\ % bad
SpeechTokenizer & 1024 & 1 & 50 & 500 & 0.63 & 1.31 & 1.14 & 1.27 & 0.17 & 7.67 \\
X-codec & 1024 & 1 & 50 & 500 & 0.84 & 2.22 & 1.71 & 3.84 & 0.49 & 3.48 \\
WavTokenizer & 4096 & 1 & 40 & 480 & 0.85 & 2.06 & 1.62 & 3.57 & 0.48 & 10.88 \\
% Single-Codec & 8192 & 1 & 23.4 & 304 &  &  &  &  &  &  \\
\midrule
SAC (ours) & 16384/16384 & 1/1 & 37.5 & 525 & \textbf{0.90} & \textbf{2.74} & \textbf{2.18} & \textbf{4.27} & \textbf{0.78} & \textbf{2.53} \\
\bottomrule
\end{tabular}
}
\caption{Comparison of low-bitrate codec models on speech reconstruction metrics.}
\label{tab:codec-low-recon}
\vspace{-2mm}
\end{table*}

\paragraph{Speech Representation}\hspace{-1em}
We evaluate the semantic richness of codec representations using the speech domain of the ARCH benchmark, following~\citet{wavtokenizer} and~\citet{unicodec}. ARCH includes RAVDESS~\citep{RAVDESS} and EMOVO~\citep{EMOVO} for emotion recognition, SLURP~\citep{SLURP} for intent classification, and AudioMNIST~\citep{Audiomnist} for digit recognition. For each codec model, the quantized representations extracted from the codec quantizer are average-pooled over time and passed through a linear classifier, following the standard ARCH protocol. For semantic–acoustic decoupled codecs such as SemantiCodec and SAC, features from the two streams are concatenated before linear probing. To further contextualize the performance gap between discrete codec representations and continuous SSL features, we additionally report results for wav2vec 2.0~\citep{wav2vec2}, data2vec~\citep{data2vec}, HuBERT~\citep{hubert}, and WavLM~\citep{wavlm} in ARCH as SSL-based references.

\begin{table*}[t]
\centering
\small
% \resizebox{1\linewidth}{!}{
\begin{tabular}{l!{\vrule width 0.8pt}lccccccc}
\toprule
Category & Model & Token Rate & BPS & RAVDESS$\uparrow$ & EMOVO$\uparrow$ & SLURP$\uparrow$ & AM$\uparrow$ & Avg.$\uparrow$ \\
\midrule
\multirow{4}{*}{\textit{\textcolor{gray}{SSL Models}}} 
& \textcolor{gray}{wav2vec 2.0} & \textcolor{gray}{-} & \textcolor{gray}{-} & \textcolor{gray}{55.32} & \textcolor{gray}{31.80} & \textcolor{gray}{14.37} & \textcolor{gray}{86.38} & \textcolor{gray}{46.97} \\
& \textcolor{gray}{data2vec} & \textcolor{gray}{-} & \textcolor{gray}{-} & \textcolor{gray}{48.03} & \textcolor{gray}{27.27} & \textcolor{gray}{\textbf{43.57}} & \textcolor{gray}{99.06} & \textcolor{gray}{54.48} \\
& \textcolor{gray}{Hubert} & \textcolor{gray}{-} & \textcolor{gray}{-} & \textcolor{gray}{\uline{65.28}} & \textcolor{gray}{\uline{40.48}} & \textcolor{gray}{\uline{33.75}} & \textcolor{gray}{\textbf{99.58}} & \textcolor{gray}{\uline{59.77}} \\
& \textcolor{gray}{WavLM} & \textcolor{gray}{-} & \textcolor{gray}{-} & \textcolor{gray}{\textbf{67.94}} & \textcolor{gray}{\textbf{43.08}} & \textcolor{gray}{30.98} & \textcolor{gray}{\uline{99.50}} & \textcolor{gray}{\textbf{60.38}} \\
\midrule
\multirow{11}{*}{\textit{Codec Models}} 
& Encodec$^\dagger$ & 150 & 1500 &  27.43 & 21.93 & 6.27 & 36.49 & 23.03 \\
& SemantiCodec & 100 & 1400 &  44.79 & 26.87 & 15.35 & 98.19 & 46.30 \\
& BigCodec & 80 & 1040 &  34.72 & 17.52 & 7.72 & 65.66 & 31.40 \\
& DAC$^\dagger$ & 100 & 1000 &  25.00 & 22.78 & 7.13 & 62.87 & 29.45 \\
& XY-Tokenizer & 100 & 1000 &  48.96 & 24.66 & 17.98 & 96.22 & 46.96 \\
& WavTokenizer$^\dagger$ & 75 & 900 & 32.55 & 31.63 & 8.02 & 69.57 & 35.44 \\
& MagiCodec & 50 & 850 &  32.99 & 25.17 & 7.73 & 70.11 & 34.00 \\
& X-Codec2 & 50 & 800 &  37.15 & 22.11 & 7.71 & 68.59 & 33.89 \\
% \midrule
& SAC & 62.5 & 875 &  \uline{57.99} & \textbf{40.31} & \textbf{29.94} & \uline{99.52} & \uline{56.94} \\
& SAC & 37.5 & 525 &  \textbf{61.81} & \uline{39.63} & \uline{29.21} & \textbf{99.63} & \textbf{57.57} \\
% SAC$_{\text{\textit{sem}}}$ & 175 & -- & -- & -- & -- & -- \\
& \textcolor{gray}{SAC$_{\text{sem}}$} & \textcolor{gray}{12.5} & \textcolor{gray}{175} &  \textcolor{gray}{39.93} & \textcolor{gray}{32.82} & \textcolor{gray}{29.37} & \textcolor{gray}{99.80} & \textcolor{gray}{50.48} \\
\bottomrule
\end{tabular}
% }
\caption{Semantic representation evaluation on the speech domain of the ARCH benchmark. The best results in each category are in \textbf{bold}, and the second-best are \uline{underlined}. \textcolor{gray}{SAC$_{\text{sem}}$} denotes evaluation using representations extracted solely from the semantic tokenizer of SAC. $^\dagger$Results are cited from the Wavtokenizer~\citep{wavtokenizer} paper.}
\label{tab:representation}
\vspace{-2mm}
\end{table*}

\section{Experimental Results and Discussions}
% \vspace{-2mm}
\subsection{Speech Reconstruction Results}
Tables~\ref{tab:codec-high-recon} and \ref{tab:codec-low-recon} present the reconstruction performance of SAC compared with existing neural audio codecs under different bitrate settings.  

At high bitrates, SAC achieves state-of-the-art performance in acoustic quality, intelligibility, and speaker similarity compared to models with similar bitrates. For reconstruction-oriented metrics such as STOI and PESQ, SAC significantly outperforms models below 1.5 kbps, while performing only slightly worse than BigCodec, which benefits from a higher token rate of 80 Hz. Notably, SAC attains a WER of 2.35\%, which is very close to the ground truth (2.16\%). We attribute this to the dual-stream architecture, which effectively disentangles semantic and acoustic modeling, enabling the semantic stream to preserve linguistic content with high fidelity. SAC also achieves a UTMOS score of 4.25, surpassing all other models as well as the ground truth. We hypothesize that this improvement stems from the acoustic stream being unconstrained from semantic modeling objectives, thereby better capturing fine-grained acoustic details (see Section~\ref{sec:speech-decoupling} for further discussion).  

At low bitrates, SAC likewise delivers state-of-the-art performance across all metrics, demonstrating the robustness of the framework under varying bandwidths. Remarkably, at a token rate of only 37.5, SAC achieves a SIM score of 0.78, exceeding the second-best model by 0.15. Moreover, its WER (2.53\%) and UTMOS (4.27) remain comparable and even surpass those of the high-bitrate setting, confirming that the dual-stream design remains effective at low bitrates: the semantic stream preserves intelligibility, while the acoustic stream complements and enhances acoustic detail. 

We include a Mean Opinion Score (MOS) \textbf{subjective evaluation} to assess the perceptual quality of the reconstructed audio, with details in Appendix~\ref{app:subjective_eval}.
To further examine robustness, we also evaluate SAC’s reconstruction performance \textbf{under noisy conditions}, as presented in Appendix~\ref{app:SAC_recon_noisy}.

\subsection{Representations Evaluation Results}
Table~\ref{tab:representation} compares SAC with other models on the semantic representation tasks using the ARCH benchmark. SAC achieves performance that substantially surpasses existing codec models, exceeding the second-best XY-Tokenizer by roughly 10\% in overall accuracy. Remarkably, SAC’s semantic representations even outperform some SSL models such as wav2vec 2.0 and data2vec, while remaining competitive with HuBERT and WavLM.  

Several findings emerge from this evaluation: \\
(1) Codecs trained without semantic supervision exhibit weaker semantic representation ability, as observed in Encodec, BigCodec, and DAC;
(2) Freezing the semantic stream during training and disentangling it from the reconstruction objective leads to significantly stronger semantic representations, as demonstrated by SemantiCodec, XY-Tokenizer, and SAC;
(3) SAC’s semantic tokenizer shows strong text-related representation ability, yielding superior performance on intent and digit classification tasks such as SLURP and AudioMNIST. Meanwhile, the acoustic stream effectively complements para-linguistic representation, leading to large gains on emotion recognition tasks (e.g., SAC outperforms SAC$_{\text{sem}}$ by about 20\% accuracy on RAVDESS);
(4) SAC delivers consistent representation performance across different bitrates. We conjecture that more fine-grained downstream tasks may be needed to investigate differences, since the current evaluation relies on global pooling.

\subsection{Effect of Auxiliary Feature Supervision}
The effects of speaker and semantic feature supervision on SAC are examined through ablation studies. As shown in Table~\ref{tab:aux-feature-ablation}, removing speaker supervision yields a slight improvement in PESQ but causes a sharp drop in SIM, from 0.78 to 0.65. This demonstrates that the proposed speaker feature supervision effectively guides the codec to preserve timbre information during training, with only a minor trade-off in reconstruction fidelity. 

When semantic supervision is removed, the model shows only a slight drop in PESQ, with other metrics remaining unaffected. This contrasts with “X-shaped” codec models, such as XY-Tokenizer, which suffer substantial degradation in ASR probing without semantic supervision. We attribute this robustness to SAC’s dual-stream design, which explicitly disentangles the semantic stream from acoustic reconstruction, thereby preserving linguistic content without relying heavily on additional supervision. In contrast, “X-shaped” models fuse semantic and acoustic streams before quantization, resulting in interference between the two objectives and making explicit semantic supervision crucial.

The impact of $\mathcal{L}_{\text{spk}}$ and $\mathcal{L}_{\text{sem}}$ on semantic representation is further analyzed in Appendix~\ref{app:ablation_feature_semantic}.

\begin{table}[htbp]
\centering
\small
\resizebox{1\linewidth}{!}{
\begin{tabular}{l!{\vrule width 0.8pt}cccccc}
\toprule
Model & STOI$\uparrow$ & \makecell{PESQ \\ NB$\uparrow$} & \makecell{PESQ \\ WB$\uparrow$} & UTMOS$\uparrow$ & SIM$\uparrow$ & WER(\%)$\downarrow$ \\
\midrule
SAC & \textbf{0.90} & 2.74 & 2.18 & 4.27 & \textbf{0.78} & \textbf{2.53}  \\
\hspace{1em}w/o $\mathcal{L}_{\text{spk}}$ & \textbf{0.90} & \textbf{2.76} & \textbf{2.21} & \textbf{4.28} & 0.65 & \textbf{2.53} \\ % speaker supervision 
\hspace{1em}w/o $\mathcal{L}_{\text{sem}}$ & \textbf{0.90} & 2.72 & 2.17 & 4.27 & \textbf{0.78} & \textbf{2.53} \\ % semantic supervision 
\bottomrule
\end{tabular}
}

\caption{Ablation study of auxiliary feature supervision.}
\label{tab:aux-feature-ablation}
% \vspace{-3mm}
\end{table}

\subsection{Downstream Speech-LLM Performance}
\label{sec:slm_performance}
We conduct downstream LLM-based experiments to assess the transferability of SAC when integrated with speech language models. Given SAC’s strong semantic representations and its previously verified ASR performance~\citep{glm-4-voice}, our focus here is on its effectiveness in generative tasks. We adopt a single-stage autoregressive (AR) TTS framework using the pre-trained Qwen3-0.6B~\citep{qwen3} as the backbone, and accommodate SAC’s dual-stream tokens through an interleaved flattening scheme based on their token-rate ratio. Using the 37.5 Hz SAC tokenizer, we train the TTS model on a 100k-hour bilingual (Chinese–English) corpus. 
As shown in Table~\ref{tab:tts_result}, the resulting system substantially outperforms state-of-the-art pure AR models, including Spark-TTS~\citep{spark-tts} and Llasa~\citep{llasa-xcodec2}, achieving significantly lower WER and higher UTMOS, which reflects superior semantic clarity and objective perceptual quality. Although speaker similarity exhibits a slight drop, we attribute this to the low-bitrate SAC variant used under current computational constraints. Detailed setup and analyses are provided in Appendix~\ref{app:tts}.

\begin{figure*}[t]
    \centering
    \includegraphics[width=\textwidth]{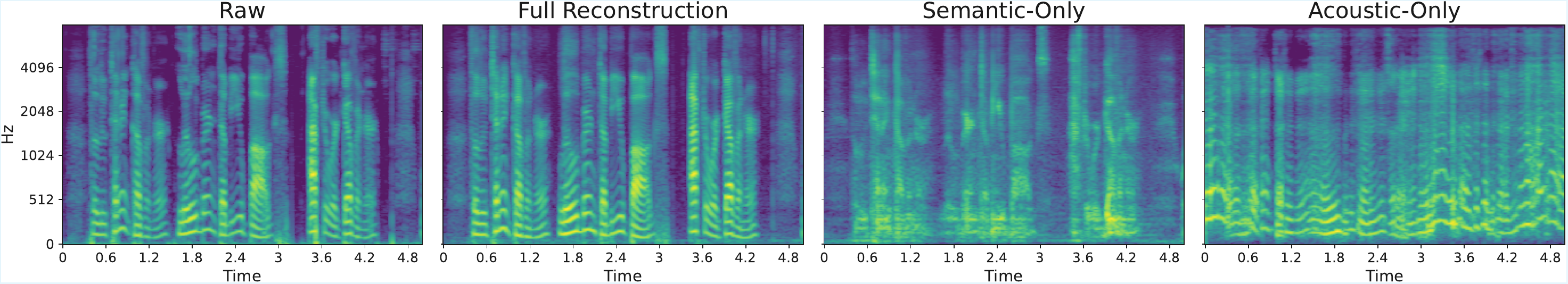}
    \caption{Mel-spectrograms of the original audio and SAC reconstructions under different reconstruction patterns.}
    \label{fig:mel_analysis}
    % \vspace{-2mm}
\end{figure*}

\begin{table}[htbp]
\centering
\small
\resizebox{1\linewidth}{!}{
\begin{tabular}{c!{\vrule width 0.8pt}lcccc}
\toprule
\makecell{Reconstruction \\ Pattern} & Model & BPS & WER$^\dagger$(\%) & SIM & MSIM \\
\midrule
\multirow{2}{*}{\textit{Full}} 
& SemantiCodec & 1400 & 3.25 & 0.72 & - \\
& SAC & 525 & 2.77 & 0.78 & - \\
\midrule
\multirow{2}{*}{\textit{Semantic-Only}} 
& SemantiCodec & 750 & 30.67 & 0.31 & 0.29 \\
& SAC & 175 & 3.99 & 0.17 & 0.64 \\
\bottomrule
\end{tabular}
}
\caption{Comparison of speech information disentanglement. $^\dagger$WER is evaluated using whisper-large-v3 for more robust and accurate speech recognition.}
\label{tab:disentanglement}
\vspace{-3mm}
\end{table}

\subsection{Speech Decoupling Analysis}
\label{sec:speech-decoupling}
In terms of speech disentanglement, we compare SAC with another semantic–acoustic decoupled codec, SemantiCodec, under the \textit{semantic-only} reconstruction pattern, where acoustic features are masked out and only the semantic stream is used for decoding.
As shown in Table~\ref{tab:disentanglement}, SAC achieves a WER of 3.99, far lower than SemantiCodec’s 30.67. 
This demonstrates that SAC’s semantic stream effectively preserves linguistic content with minimal interference from acoustic representations during decoding.
We further evaluate SIM and mean similarity (MSIM) for \textit{semantic-only} reconstructions. MSIM is computed as the average cosine similarity between speaker embeddings across all reconstructed utterances. Results show that SAC’s SIM is only 0.17, demonstrating a clean disentanglement of timbre information from the original audio. Meanwhile, its MSIM reaches 0.64, indicating that \textit{semantic-only} reconstructions converge toward a uniform timbre, with subjective listening revealing a consistent male bass voice. These findings highlight SAC’s superior timbre disentanglement and suggest potential speech applications such as speaker anonymization.

As illustrated in Fig.~\ref{fig:mel_analysis}, we performed a visual analysis of SAC’s reconstruction capabilities across different paradigms. In the \textit{full} reconstruction setting, which follows SAC's basic reconstruction method using dual-stream tokens, the spectrogram preserves more high-frequency textures compared to the original, indicating reduced distortion in harmonic structures and formants. This observation also explains SAC’s strong performance on UTMOS, as the decoder effectively acts as a generator to enrich fine-grained acoustic detail.  

In contrast, the \textit{semantic-only} reconstruction lacks the speaker-specific fundamental frequencies and formants, yet retains clear semantic content. This demonstrates SAC’s ability to effectively separate and preserve the semantic features, while discarding speaker-dependent acoustic information. On the other hand, the \textit{acoustic-only} reconstruction retains well-defined fundamental frequencies and formants, but completely loses semantic content (The output is perceptually similar to noise, devoid of any meaningful information). This highlights SAC’s remarkable capacity to disentangle and independently process semantic and acoustic features, providing direct evidence of its robust disentanglement ability. Additional mel-spectrogram reconstruction analyses of other codec models are provided in Appendix~\ref{app:mel_compare}.

\section{Conclusion}
In this paper, we presented SAC, a neural speech codec with semantic–acoustic dual-stream quantization. To exploit the strengths of semantic tokenizers for capturing linguistic content and codecs for modeling acoustic detail, SAC introduces independent semantic and acoustic streams that extract tokens separately. To further enhance timbre modeling, we incorporated speaker feature supervision into codec training.  
Comprehensive evaluations demonstrate that SAC achieves SOTA performance in both speech reconstruction and semantic representation across different bitrates, while ablation studies validate the effectiveness of auxiliary feature supervision. 
Moreover, downstream LLM-based TTS experiments further confirm SAC’s effectiveness as a speech tokenizer for generative speech applications.
We also observe a remarkably clean disentanglement between semantic and acoustic tokens in reconstruction: semantic-token-based reconstruction contains no speaker-related information, while acoustic-token-based reconstruction preserves no semantic content. To the best of our knowledge, this is the first instance of such a clean disentanglement between semantics and speaker identity in terms of reconstruction. This finding offers new insights for subsequent tasks that rely on disentanglement, such as voice conversion or the joint control of style and timbre in TTS.

% \newpage
\section*{Limitations}
Although SAC demonstrates superior reconstruction quality and semantic representation within the speech domain, its generalizability to other types of audio signals, such as music and sounds, remains to be explored. A key challenge lies in the semantic tokenizer used in SAC, which is trained under ASR supervision on speech data and therefore primarily aligned with textual objectives. In contrast, the semantics in music and sound extend beyond linguistic alignment. To develop a more general-purpose audio codec based on the SAC dual-stream architecture, future work should focus on designing a universal audio semantic encoder, potentially through multi-task supervision across modalities or self-supervised training on diverse audio data.

% In addition, while SAC introduces speaker feature supervision to improve timbre preservation, it does not explicitly incorporate supervision for other speech attributes such as accent, emotion, or paralinguistic cues. Exploring how to integrate such attribute-aware supervision into codec training without compromising reconstruction quality represents a valuable research direction.  

\bibliography{custom}

% \newpage
% \thispagestyle{empty}
% \mbox{}
\newpage

\appendix

\section{Details of Training Data}
\label{app:SAC_data}
As summarized in Table~\ref{tab:codec-data-statistics}, SAC is trained on approximately 20,000 hours of bilingual (Chinese and English) speech, comprising around 8 million utterances. The training corpus includes roughly 10,000 hours in each language. All audio is resampled to 16 kHz. 

For Chinese, the data sources include the Chinese subset of Emilia~\citep{emilia}, WenetSpeech4TTS~\citep{wenetspeech4tts}, and a small amount of in-house para-linguistic data. For English, we use LibriSpeech~\citep{librispeech}, the small and medium subsets of Libriheavy~\citep{libriheavy}, the English subset of Emilia, MLS~\citep{mls}, and a small amount of in-house data.  

To promote diversity, we follow the annotation scheme of VoxBox~\citep{spark-tts}, which labels speech samples along three dimensions: \textit{age}, \textit{gender}, and \textit{emotion}. Specifically, we partition Emilia, WenetSpeech4TTS, and MLS into categories defined by the unique combination of these three attributes, and then sample data to achieve balanced coverage across categories. This strategy ensures broad diversity within the training set.  

All data are drawn from the official training splits of each dataset to maintain fair evaluation. Given the small scale and similarity of the in-house data to existing open-source corpora, we believe that reproducing SAC with only open-source data would not significantly affect performance.

\begin{table}[htbp]
\centering
\small
\setlength\tabcolsep{3pt}
\resizebox{1\linewidth}{!}{
\begin{tabular}{l!{\vrule width 0.8pt}crrr}
\toprule
Dataset & Lang. & \#Utt. & Dur. (h) & Avg. (s) \\
\midrule
Emilia-ZH & ZH & 3.0M & 6712 & 8.05 \\
WenetSpeech4TTS & ZH & 2.0M & 2754 & 4.96 \\
In-house Data & ZH & 0.2M & 525 & 7.82 \\
\midrule
LibriSpeech & EN & 0.3M & 961 & 12.30 \\
Libriheavy & EN & 1.2M & 5042 & 14.85 \\
Emilia-EN & EN & 1.0M & 2481 & 8.93 \\
MLS-EN & EN & 0.3M & 1222 & 14.66 \\
In-house Data & EN & 0.1M & 209 & 6.42 \\
\midrule
Summary (All) & ZH\&EN & 8.2M & 19906 & 8.78 \\
\bottomrule
\end{tabular}
}
\caption{Data statistics for SAC training.
``\#Utt.'' refers to the total number of audio samples, 
``Dur. (h)'' denotes the total duration in hours, 
and ``Avg. (s)'' indicates the average duration per sample.}
\label{tab:codec-data-statistics}
\end{table}

\begin{table*}[htbp]
\centering
\small
\resizebox{1\linewidth}{!}{
\begin{tabular}{l!{\vrule width 0.8pt}cccc!{\vrule width 0.8pt}ccccccc}
\toprule
Model & \makecell{Codebook \\ Size} & Nq & \makecell{Token \\ Rate} & BPS & STOI$\uparrow$ & \makecell{PESQ \\ NB$\uparrow$} & \makecell{PESQ \\ WB$\uparrow$} & UTMOS$\uparrow$ & SIM$\uparrow$ & WER(\%)$\downarrow$ \\
\midrule
Ground Truth & - & - & - & - & 1.00 & 4.55 & 4.64 & 3.50 & 1.00 & 4.59 \\
\midrule
X-codec & 1024 & 2 & 100 & 1000 & 0.85 & 2.50 & 1.99 & 3.66 & 0.67 & 7.02 \\
XY-Tokenizer & 1024 & 8 & 100 & 1000 & 0.89 & 2.80 & 2.23 & 3.46 & 0.82 & 6.19 \\
WavTokenizer & 4096 & 1 & 75 & 900 & 0.87 & 2.40 & 1.96 & 3.22 & 0.68 & 13.35 \\
MagiCodec & 131072 & 1 & 50 & 850 & \textbf{0.90} & \textbf{2.94} & 2.34 & 3.70 & 0.75 & 10.63 \\
X-codec2 & 65536 & 1 & 50 & 800 & \textbf{0.90} & 2.83 & 2.26 & 3.64 & 0.81 & 6.85 \\
\midrule
SAC (ours) & 16384/16384 & 1/1 & 62.5 & 875 & \textbf{0.90} & 2.92 & \textbf{2.39} & \textbf{3.84} & \textbf{0.85} & \textbf{5.77} \\
\bottomrule
\end{tabular}
}
\caption{Comparison of high-bitrate codecs on speech reconstruction metrics under noisy conditions.}
\label{tab:codec-high-recon-noisy}
\end{table*}

\begin{table*}[ht]
\centering
\small
\resizebox{1\linewidth}{!}{
\begin{tabular}{l!{\vrule width 0.8pt}cccc!{\vrule width 0.8pt}ccccccc}
\toprule
Model & \makecell{Codebook \\ Size} & Nq & \makecell{Token \\ Rate} & BPS & STOI$\uparrow$ & \makecell{PESQ \\ NB$\uparrow$} & \makecell{PESQ \\ WB$\uparrow$} & UTMOS$\uparrow$ & SIM$\uparrow$ & WER(\%)$\downarrow$ \\
\midrule
Ground Truth & - & - & - & - & 1.00 & 4.55 & 4.64 & 3.50 & 1.00 & 4.59 \\
\midrule
SpeechTokenizer & 1024 & 1 & 50 & 500 & 0.61 & 1.27 & 1.12 & 1.27 & 0.15 & 19.85 \\
X-codec & 1024 & 1 & 50 & 500 & 0.82 & 2.10 & 1.63 & 3.47 & 0.48 & 9.58 \\
WavTokenizer & 4096 & 1 & 40 & 480 & 0.82 & 1.95 & 1.56 & 3.16 & 0.51 & 30.28 \\
\midrule
SAC (ours) & 16384/16384 & 1/1 & 37.5 & 525 & \textbf{0.87} & \textbf{2.54} & \textbf{2.03} & \textbf{3.90} & \textbf{0.77} & \textbf{6.36} \\
\bottomrule
\end{tabular}
}
\caption{Comparison of low-bitrate codecs on speech reconstruction metrics under noisy conditions.}
\label{tab:codec-low-recon-noisy}
\end{table*}

\section{Codec Training Details}
\label{app:SAC_train_details}
To accelerate SAC training, we pre-extracted all semantic representations $\mathbf{S}_c$ and corresponding semantic tokens from the training corpus using a pretrained semantic tokenizer \citep{glm4voice-tokenizer}. During training, only the codebook from the tokenizer is required to obtain the quantized semantic embeddings $\mathbf{S}_q$, while $\mathbf{S}_c$ serves as the ground-truth target for auxiliary semantic feature supervision. The generator contains approximately 277M parameters, with around 249M being trainable. The codebook of the semantic tokenizer and the speaker encoder are frozen to ensure that well-pretrained semantic and speaker features are properly utilized.  

For generator training, the loss coefficients are set as $\lambda_{\text{recon}} = 15$, $\lambda_{\text{vq}} = 1$, $\lambda_{\text{adv}} = 1$, $\lambda_{\text{feat}} = 2$, $\lambda_{\text{sem}} = 1000$, and $\lambda_{\text{spk}} = 10$. Within the VQ loss, the commitment and codebook loss weights are set to 0.25 and 4, respectively. To improve training stability, the generator is pretrained for 1,500 steps before introducing the discriminator for adversarial training. An exponential moving average (EMA) is applied to maintain a smoothed version of the model parameters \citep{AUV}, which are used during inference and observed to enhance model stability.

\section{Speech Reconstruction under Noisy Conditions}
\label{app:SAC_recon_noisy}
To further evaluate the robustness of SAC in noisy environments, we conduct additional reconstruction experiments on the full \textit{LibriSpeech test-other} set, which contains significantly more background noise compared to the \textit{test-clean} set. Tables~\ref{tab:codec-high-recon-noisy} and~\ref{tab:codec-low-recon-noisy} present the reconstruction results of SAC under high-bitrate and low-bitrate settings, respectively, alongside comparable state-of-the-art codecs.

As expected, compared to the results on the \textit{test-clean} set, the ground truth recordings exhibit notable degradation in noisy conditions, with UTMOS decreasing from 4.09 to 3.50 and WER increasing from 2.16\% to 4.59\%. All codecs show a general decline in reconstruction quality, as reflected by reduced STOI, PESQ-NB, and PESQ-WB scores. Nevertheless, SAC consistently achieves the best performance across bitrates, closely mirroring its relative ranking in the \textit{test-clean} setting. 

Notably, SAC maintains a clear advantage in UTMOS, SIM, and WER.  
For objective naturalness, SAC achieves UTMOS scores of 3.84 and 3.90 at high and low bitrates, respectively—both substantially higher than the ground truth value of 3.50. We attribute this to SAC’s acoustic stream effectively modeling fine-grained acoustic details, while the decoder functions as a generator that enhances objective perceptual realism. In terms of speaker similarity, SAC shows only a marginal drop of 0.01 in SIM compared to the clean condition, whereas other codecs experience significant declines. This demonstrates the robustness of both our model architecture and the diverse training data, as well as the effectiveness of the speaker feature supervision in preserving timbre characteristics.  
Furthermore, SAC maintains high speech intelligibility under noisy conditions, with remarkably low WER values, confirming that its dual-stream architecture enables reliable semantic reconstruction even in the presence of noise. 

These findings highlight SAC’s robustness under extreme high compression and its ability to deliver high-fidelity speech reconstruction in both clean and noisy environments—underscoring its strong potential for applications in speech compression and transmission.

\section{Downstream LLM-Based Speech Generation}
\label{app:tts}

To validate the potential of SAC in downstream speech language models (SLMs) tasks, we further conducted an LLM-based text-to-speech (TTS) experiment. 
For language modeling, we adopt a pure autoregressive (AR) framework using the pre-trained LLM Qwen3-0.6B~\footnote{\url{https://huggingface.co/Qwen/Qwen3-0.6B}}~\citep{qwen3} as the backbone. In contrast to prior TTS systems such as VALL-E~\citep{valle}, which rely on a two-stage AR+NAR generation pipeline, our approach uses a single Transformer decoder to autoregressively predict single-layer speech tokens, substantially simplifying the modeling process.

To accommodate SAC’s dual-stream tokens, we introduce a novel interleaved flattening strategy for language modeling. Since SAC produces single-layer semantic and acoustic tokens with different token rates, we arrange them in an interleaved sequence proportional to their rate ratio. For example, the 37.5 Hz SAC variant generates semantic tokens at 12.5 Hz and acoustic tokens at 25 Hz; the downstream model therefore predicts them in a fixed 1:2 pattern in a single layer, where earlier semantic predictions facilitate subsequent acoustic predictions.

Given computational constraints, we use the 37.5 Hz SAC tokenizer and expand the LLM vocabulary with both semantic and acoustic codebooks. During training, the decoder-only LLM is optimized via negative log-likelihood to predict speech tokens conditioned on text transcriptions as prefixes.

Small-scale TTS evaluations often fail to reflect the robustness and generalization abilities of speech tokens, especially when paired with LLMs whose interaction dynamics exhibit complex emergent behavior. To obtain a more reliable assessment, we train the TTS model using SAC tokens on a large 100k-hour bilingual corpus (Chinese and English), following the same data distribution as VoxBox~\citep{spark-tts}.

We evaluate the zero-shot TTS performance on the Seed-TTS-eval ~\citep{seed-tts}, using speaker similarity (SIM), WER (or CER for the Chinese test set), and UTMOS as evaluation metrics. During inference, we concatenate the prompt text, the target synthesis text, and the prompt-audio token sequence as the input to the LLM.
For a fair comparison, we benchmark our model against two state-of-the-art purely AR TTS systems trained on large-scale datasets: Spark-TTS~\citep{spark-tts} and Llasa~\citep{llasa-xcodec2}. Spark-TTS adopts BiCodec as the speech tokenizer and Qwen2.5-0.5B as the backbone, trained on the same VoxBox data. Llasa employs X-codec2 as the tokenizer and Llama 3.2-1B~\citep{Llama} as the backbone, with versions trained on 80k, 160k, and 250k hours of data.

\begin{table}[htbp]
\centering
\small
\resizebox{1\linewidth}{!}{
\renewcommand{\arraystretch}{1.2}
\begin{tabular}{lccc}
\toprule
\textbf{Model} & \textbf{WER(\%)}$\downarrow$ & \textbf{SIM}$\uparrow$ & \textbf{UTMOS}$\uparrow$ \\
\midrule
\multicolumn{4}{c}{\textbf{Seed-TTS \textit{test-en}}} \\
\midrule
Llasa-1B-80k     & 3.71 & 0.54 & 4.06 \\
Llasa-1B-160k    & 3.60 & 0.56 & 4.05 \\
Llasa-1B-250k    & 2.99 & 0.57 & 4.07 \\
Spark-TTS        & 1.98 & \textbf{0.58} & 3.94 \\
\hdashline
\textbf{Ours}    & \textbf{1.06} & 0.54 & \textbf{4.21} \\
\midrule
\multicolumn{4}{c}{\textbf{Seed-TTS \textit{test-zh}}} \\
\midrule
Llasa-1B-80k     & 2.69 & 0.65 & 3.27 \\
Llasa-1B-160k    & 2.22 & 0.66 & 3.28 \\
Llasa-1B-250k    & 1.89 & \textbf{0.67} & 3.28 \\
Spark-TTS        & 1.20 & \textbf{0.67} & 3.27 \\
\hdashline
\textbf{Ours}    & \textbf{0.90} & 0.65 & \textbf{3.34} \\
\bottomrule
\end{tabular}
}
\caption{Results of our TTS model compared with prior single-stage AR TTS systems on the Seed-TTS test sets. The evaluation results for Spark-TTS and Llasa are cited from their respective papers.}
\label{tab:tts_result}
\end{table}

Table~\ref{tab:tts_result} reports the zero-shot TTS performance of our model trained with the 37.5 Hz SAC tokenizer. Compared with prior pure AR TTS systems, our model achieves substantial gains in both intelligibility and objective naturalness. On the \textit{test-en} set, our model attains a WER of 1.06\%, markedly outperforming the  Spark-TTS (1.98\%). On the \textit{test-zh} set, we obtain a CER of 0.90\%, achieving a performance below 1\% for the first time. To the best of our knowledge, our TTS model is optimal or near-optimal in terms of semantic clarity, even when compared against all existing TTS models (including non-purely AR models). Moreover, our model yields consistently higher UTMOS scores than both Llasa and Spark-TTS, demonstrating superior objective naturalness. These results collectively highlight the promising potential of SAC in downstream generative tasks and further confirm its capacity for high-fidelity semantic preservation and high-naturalness speech synthesis.

Despite these improvements, our model shows a slight reduction in speaker similarity compared with previous AR systems. We attribute this mainly to the low token rate of the 37.5 Hz SAC tokenizer: its coarse temporal granularity limits the capacity to capture fine-grained timbre details relative to the 50 Hz codecs used in prior TTS models. 

Nevertheless, we believe that SAC has already demonstrated strong potential for generative downstream tasks. Moreover, for applications involving semantic or acoustic editing, the semantic–acoustic decoupled token design offers a particularly promising modeling foundation. As future work, we plan to investigate how data scaling and model scaling influence generation quality—especially speaker similarity—and to evaluate higher-rate SAC variants (e.g., 62.5 Hz) in zero-shot TTS, which remains unexplored due to current computational limitations.

\section{Subjective Evaluation on Speech Reconstruction}
\label{app:subjective_eval}
Since reconstructed audio from different codec models at high bitrates (particularly above 800 bps) tends to be perceptually indistinguishable to human listeners, we focus our subjective evaluation on low-bitrate settings. In this regime, Encodec~\citep{encodec} and SpeechTokenizer~\citep{speechtokenizer} produce notably low reconstruction quality when restricted to their first-layer RVQ tokens (with STOI scores below 0.80), and TS3-Codec~\citep{Ts3-codec} does not provide publicly available model weights. Consequently, these models are excluded from our comparison. Our Mean Opinion Score (MOS) study therefore includes Ground Truth, SemanticCodec~\citep{semanticodec}, X-codec~\citep{xcodec}, WavTokenizer~\citep{wavtokenizer}, and SAC (with a token rate of 37.5 Hz).

For the evaluation, 20 native speakers were invited to assess reconstructed audio samples generated by each model. A total of 30 utterances were randomly selected from the LibriSpeech \textit{test-clean} set. Evaluators were thoroughly informed of the scoring criteria and instructed to judge naturalness and perceptual quality as the primary factors in their evaluation. Each sample was rated on a 1–5 scale with 0.5-point increments, where higher scores indicate better perceived quality.

\begin{table}[t]
\centering
\small
\begin{tabular}{l !{\vrule width 0.8pt} c c c}
\toprule
Model & Token Rate & BPS & MOS$\uparrow$ \\
\midrule
Ground Truth & - & - & 3.98 \\
\midrule
SemantiCodec & 50 & 700 & 2.80 \\ 
X-codec & 50 & 500 & 3.18 \\ 
WavTokenizer & 40 & 480 & 3.20 \\ 
\midrule
SAC (ours) & 37.5 & 525 & \textbf{3.94} \\
\bottomrule
\end{tabular}
\caption{Subjective evaluation of reconstructed audio from different codec models on low-bitrate settings.}
\label{tab:subjective_codecs}
\end{table}

As shown in Table~\ref {tab:subjective_codecs}, SAC achieves a MOS of 3.94, substantially higher than all other codec baselines and approaching the ground-truth score of 3.98. This indicates that SAC preserves perceptual quality with almost no audible degradation, whereas competing codecs introduce noticeable distortion. The subjective results also highlight the limitations of UTMOS as a model-based predictor: although it reliably captures overall quality trends, finer-grained distinctions still require human evaluation. For example, WavTokenizer receives a lower UTMOS score than X-Codec (3.57 vs.~3.84), yet is slightly preferred in the human MOS assessment.

\section{Impact of Training Data Scale}
\label{app:data_scaling}
Existing codec models are often trained on datasets of varying scales, which complicates a fair comparison of their performance. For example, WavTokenizer~\citep{wavtokenizer} is trained on 8k hours of mixed data, XY-Tokenizer~\citep{XY-Tokenizer} utilizes 101k hours of the multilingual Emilia~\citep{emilia} dataset, and TS3-Codec~\citep{Ts3-codec} uses 60k hours of Libri-light~\citep{libri-light} data.
To validate the generality and scalability of the SAC modeling approach, we further include results trained exclusively on the LibriSpeech~\citep{librispeech} 960-hour English speech corpus (based on the low-bitrate version of SAC). This represents the smallest training data volume among all comparable bitrate codecs.

\begin{table}[h]
\centering
\small
\resizebox{1\linewidth}{!}{
\begin{tabular}{l!{\vrule width 0.8pt}cccccc}
\toprule
Model & STOI$\uparrow$ & \makecell{PESQ \\ NB$\uparrow$} & \makecell{PESQ \\ WB$\uparrow$} & UTMOS$\uparrow$ & SIM$\uparrow$ & WER(\%)$\downarrow$ \\
\midrule
$\text{SAC}_{large}$ & \textbf{0.90} & \textbf{2.74} & \textbf{2.18} & 4.27 & \textbf{0.78} & \textbf{2.53} \\
$\text{SAC}_{small}$ & \textbf{0.90} & 2.70 & 2.16 & \textbf{4.30} & 0.63 & 2.58 \\ 
\bottomrule
\end{tabular}
}

\caption{Reconstruction evaluation of SAC trained with different data scales. $\text{SAC}_{large}$ denotes the model trained on the 20k hours of large-scale data mentioned in the main paper, while $\text{SAC}_{small}$ represents the model trained exclusively on the LibriSpeech dataset.}
\label{tab:data-scaling}
\end{table}

As shown in Table~\ref{tab:data-scaling}, $\text{SAC}_{small}$, trained on a small-scale dataset, still maintains superior reconstruction performance. Although its performance slightly declines in metrics such as PESQ and WER compared to $\text{SAC}_{large}$, it still demonstrates a significant advantage over other comparable-bitrate codecs presented in Table \ref{tab:codec-low-recon}. Notably, $\text{SAC}_{small}$ achieves a further improvement in UTMOS, reaching up to 4.30. We primarily attribute this to the high purity of the audio in the LibriSpeech dataset, which enables the model to achieve better reconstruction quality on the equally clean test set (LibriSpeech \textit{test-clean}) than $\text{SAC}_{large}$.

However, $\text{SAC}_{small}$ exhibits a substantial drop in SIM compared to $\text{SAC}_{large}$, decreasing from 0.78 to 0.63. We hypothesize that this is primarily caused by the severe lack of speaker diversity in the training data, leading SAC to overfit to speaker-specific characteristics. Specifically, we randomly sampled 2620 samples—the same number as the LibriSpeech test-clean set—from the training sets of both models and tested their reconstruction performance. The results show that the SIM score of $\text{SAC}_{small}$ on its training examples reached 0.72, which is significantly higher than 0.63 on the test set. In contrast, the SIM score of $\text{SAC}_{large}$ on its training examples remained consistent with the test examples (both 0.78), fully illustrating the robustness and generalization capability provided by large-scale training data.

In summary, this analysis indicates that SAC can maintain outstanding reconstruction performance even in low-resource scenarios, demonstrating excellent semantic clarity and acoustic quality. Furthermore, scaling up data, particularly by increasing the speaker diversity of the training corpus, effectively mitigates the issue of model overfitting to speaker characteristics, which further validates the generalization and scalability of the SAC model.

\section{Speech Reconstruction Speed}
\label{app:rtf}
To evaluate the real-time performance of different speech codec models in speech reconstruction, we conduct experiments that measure the Real-Time Factor (RTF), defined as the ratio of processing time to audio duration. Specifically, 1,000 samples are randomly selected from the seed-eval dataset~\citep{seed-tts}, and reconstruction is performed on an NVIDIA L20 GPU with a batch size of 1. For comparison, SAC is evaluated alongside two speech codecs of similar model scale—SemantiCodec~\citep{semanticodec} and XY-Tokenizer~\citep{XY-Tokenizer}.

\begin{table}[htbp]
\centering
\small
\setlength\tabcolsep{3pt}
\begin{tabular}{l!{\vrule width 0.8pt}cccc}
\toprule
Models & BPS & \#Params & RTF$\downarrow$ \\
\midrule
SemantiCodec & 1400 & 507M & 0.3608  \\
XY-Tokenizer & 1000 & 520M & 0.0155  \\
SAC & 875 & 533M & \textbf{0.0135}  \\
\bottomrule
\end{tabular}
\caption{Real-Time Factor (RTF) for audio codec models on test audio clips using an L20 GPU.}
\label{tab:rtf-exp}
\end{table}

As shown in Figure~\ref{tab:rtf-exp}, despite containing over 500M parameters, SAC maintains superior real-time performance compared to codec models of similar size, achieving an RTF as low as 0.0135. This efficiency primarily stems from SAC’s architecture: apart from the semantic tokenizer, which employs Transformer blocks, all remaining components consist of lightweight convolutional and linear layers that significantly accelerate inference. These results demonstrate SAC’s high efficiency and practicality for real-world deployment and various downstream applications.

\section{Effect of Auxiliary Feature Supervision on Semantic Representation}
\label{app:ablation_feature_semantic}
To investigate the impact of semantic and speaker feature supervision on SAC's semantic representation capabilities, we further evaluate the semantic representation performance of SAC on ARCH under these two ablation settings.

\begin{table}[h]
\centering
\small
\resizebox{1\linewidth}{!}{
\begin{tabular}{l!{\vrule width 0.8pt}ccccc}
\toprule
Model & RAVDESS$\uparrow$ & EMOVO$\uparrow$ & SLURP$\uparrow$ & AM$\uparrow$ & Avg.$\uparrow$ \\
\midrule
SAC & 61.81 & 39.63 & \textbf{29.21} & \textbf{99.63} & \textbf{57.57} \\
\hspace{1em}w/o $\mathcal{L}_{\text{spk}}$ & 59.72 & \textbf{40.32} & 27.34 & 99.54 & 56.73 \\ % speaker supervision 
\hspace{1em}w/o $\mathcal{L}_{\text{sem}}$ & \textbf{62.19} & 38.78 & 29.14 & 99.60 & 57.43 \\ % semantic supervision 
\bottomrule
\end{tabular}
}
\caption{Ablation study of auxiliary feature supervision on semantic representations evaluations.}
\label{tab:ablation_feature_semantic}
\end{table}

As shown in Table~\ref{tab:ablation_feature_semantic}, removing the speaker feature supervision loss ($\mathcal{L}_{\text{spk}}$) or the semantic feature supervision loss ($\mathcal{L}_{\text{sem}}$) during training has a minimal impact on the final semantic representation evaluation results. Although the results fluctuate slightly across individual sub-evaluation sets, the overall score of the ablation models shows a decline of less than 1\% compared to the baseline. This suggests that the auxiliary supervision branches have a limited effect on the semantic representation capabilities learned by SAC, and the dual-stream quantization design itself is sufficient to ensure the model retains rich semantic information. 

Nevertheless, to achieve optimal semantic representation and reconstruction performance, we retained both feature supervision mechanisms in the final model design. However, for training efficiency, we also recommend removing the semantic supervision branch in resource-constrained scenarios, as this simplification only leads to a marginal decrease in both reconstruction performance and semantic representation capability.

\begin{figure*}[t]
    \centering
    \includegraphics[width=\textwidth]{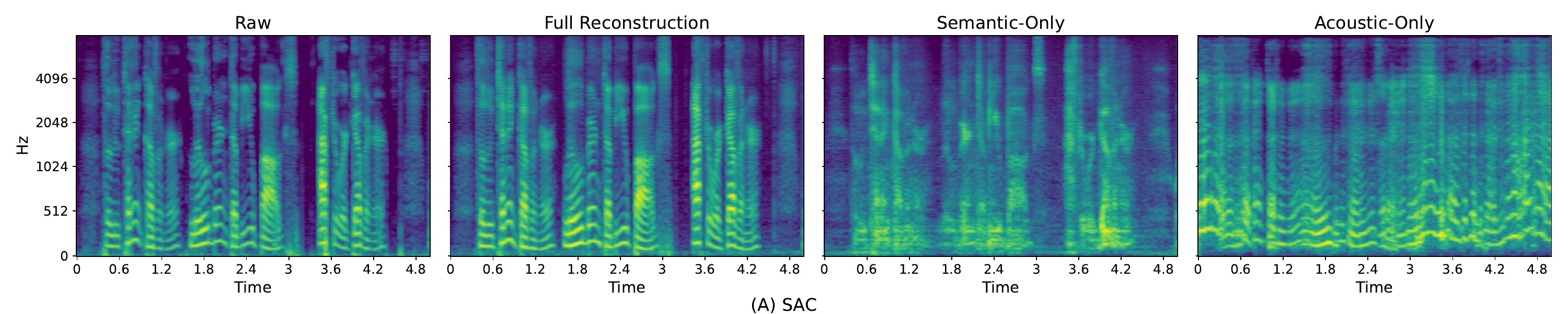}
    \includegraphics[width=\textwidth]{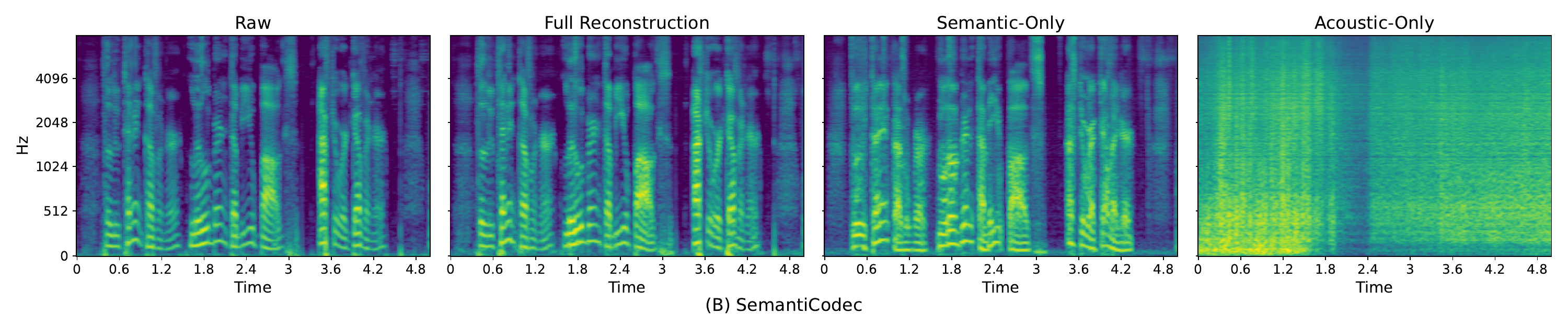} \\[1ex]
    \includegraphics[width=\textwidth]{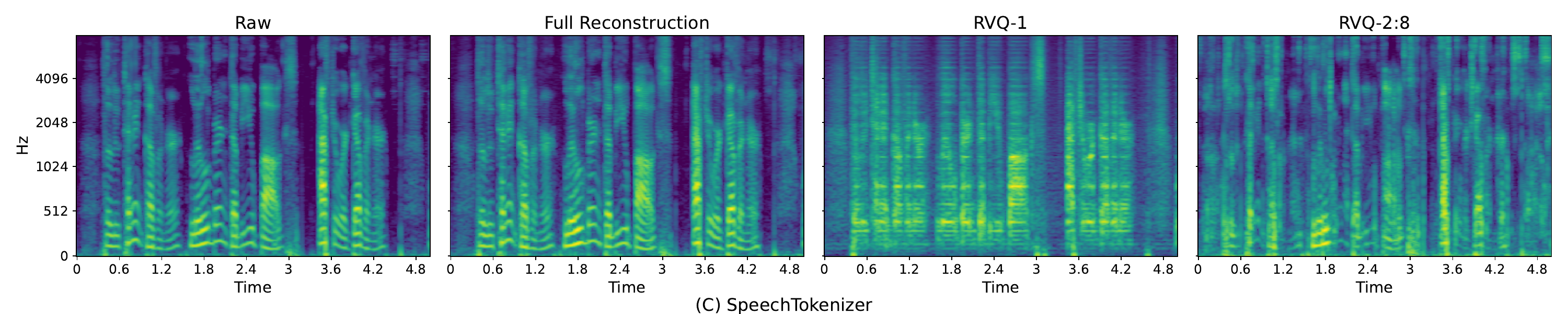} \\[1ex]
    \caption{Comparison of mel-spectrogram reconstructions from different codec models under distinct reconstruction patterns. (A) SAC, (B) SemantiCodec, and (C) SpeechTokenizer.}
    \label{fig:mel_recon_cmp}
\end{figure*}

\section{Mel-Spectrogram Reconstructions Comparison among Codecs}
\label{app:mel_compare}
To further analyze the speech disentanglement capability of different speech codecs, we visualize mel-spectrograms reconstructed under various configurations using a randomly selected speech sample from the LibriSpeech \textit{test-clean} set (identical to the one used in the main paper). In particular, we additionally include SpeechTokenizer~\citep{speechtokenizer} for comparison, which achieves effective semantic–acoustic disentanglement through its multi-layer residual vector quantization (RVQ) structure, by reconstructing speech from tokens of different RVQ layers.

In these settings, \textit{full} reconstruction refers to speech reconstructed using all discrete speech tokens. The \textit{semantic-only} reconstruction denotes cases where only semantic tokens are used, while acoustic embeddings are masked with zeros, as in semantic–acoustic decoupled codecs such as SAC and SemantiCodec~\citep{semanticodec}. In SpeechTokenizer, the first RVQ layer is guided by semantic distillation to primarily encode linguistic information; therefore, reconstruction from the first-layer tokens (denoted as \textit{RVQ-1}) can be considered analogous to \textit{semantic-only} reconstruction.
Conversely, \textit{acoustic-only} reconstruction uses only the acoustic tokens while masking semantic embeddings. In SpeechTokenizer, the modeling of acoustic details is pushed to the deeper RVQ layers (layers 2–8); thus, reconstructions based on tokens from these layers (denoted as \textit{RVQ-2:8}) correspond to the disentangled acoustic component.

Figure~\ref{fig:mel_recon_cmp} presents the reconstruction results of different speech codecs under various reconstruction settings.  
In the \textit{full} reconstruction, SAC retains a continuous harmonic structure in the mid-frequency and high-frequency regions, with clear formant shapes and spectral patterns. Notably, it even introduces richer harmonic details than the original signal, demonstrating its superior reconstruction fidelity and high UTMOS score. In contrast, SemantiCodec produces slightly blurred high-frequency components with lower spectral precision, while SpeechTokenizer reproduces the coarse spectral envelope but lacks fine texture, particularly in high-frequency regions.  

In the \textit{semantic-only} reconstruction, SAC’s harmonics almost vanish, preserving only a coarse energy contour of speech while effectively removing formant-related acoustic features. This confirms that the semantic stream in SAC is minimally influenced by the acoustic stream during decoding, achieving clean disentanglement of non-linguistic information. SemantiCodec, however, still exhibits visible harmonic and formant structures, suggesting considerable residual acoustic details. This can be attributed to its design: the semantic tokens in SemantiCodec are derived from AudioMAE~\citep{audiomae}, which is trained on general audio rather than speech data, making it less capable of separating semantic content from acoustic cues. As a result, its semantic tokens encode general informative features rather than purely linguistic representations. SpeechTokenizer’s \textit{RVQ-1} reconstruction also shows weakened low-frequency harmonics but retains high-frequency spectral density, indicating that the first RVQ layer still carries some acoustic features.  

In the \textit{acoustic-only} reconstruction, SAC preserves clear harmonic structures and temporal energy variations in the low-frequency and mid-frequency ranges, reflecting strong modeling of acoustic patterns. However, its high-frequency harmonics are relatively weaker, implying room for improvement in high-frequency modeling. SemantiCodec produces an overly smoothed spectrum with a lack of fine-grained detail, indicating limited acoustic expressiveness—likely due to the absence of semantic tokens as decoding conditions. SpeechTokenizer’s reconstruction retains fragmented and unstable harmonic structures, with fluctuating energy distributions, suggesting that its acoustic and semantic components are not cleanly separated across RVQ layers.

\end{document}